# Universality of competitive networks for weighted networks [*]

Guo Jin-Li, Fan Chao, Ji Ya-Li

Business School, University of Shanghai for Science and Technology, Shanghai, 200093

**Abstract:** In this paper, we propose a new model that allows us to investigate this competitive aspect of real networks in quantitative terms. Through theoretical analysis and numerical simulations, we find that the competitive network have the universality for a weighted network. The relation between parameters in the weighted network and the competitiveness in the competitive network is obtained. So we can use the expression of the degree distribution of the competitive model to calculate that and the strength of the weighted network directly. The analytical solution reveals that the degree distribution of the weighted network is correlated with the increment and initial value of edge weights, which is verified by numerical simulations. Moreover, the evolving pattern of a clustering coefficient along with network parameters such as the size of a network, an updating coefficient, an initial weight and the competitiveness are obtained by further simulations. Specially, it is necessary to point out that the initial weight plays equally significant role as updating coefficient in influencing the topological characteristics of the network.

## 1. Introduction

It is remarkable that, Watts et al. proposed WS model to describe the small-world effect and Barabási et al. proposed an evolving model named BA model to construct a scale-free network. They began a new era in the study of complex networks. A number of important results have emerged from this field up to now [1-5]. Due to the limitation, the BA model has been improved by a lot of scientists to describe real networks better. Bianconi and Barabási have proposed a fitness model in view of the competition phenomenon in evolving process[6], where the fitness refers to the capacity of nodes to obtain edges, so that the more competitive nodes can attract more edges. Dorogovtev et al. investigated an initial attractiveness model of directed networks, where attractiveness factor is used to describe the attraction of a new node to the old and adjust the power-law exponent [7], which the attractiveness is a constant in the model. However, different nodes usually have different attractiveness in real networks, even changing with time.

---

[*] Project supported by the National Natural Science Foundation of China (Grant Nos. 70871082) and the Shanghai First-class Academic Discipline Project (Grant No. S1201YLXK).

Both the BA model and the initial attractiveness model are more difficult to describe the mechanism of competition networks. If two countries have diplomatic relations we establish an edge between them, which form a national network of relationships. The network may not well reflect a competition of the economic and military strength. The comprehensive national strength reflects a country's competitiveness. In the course of evolution of the national network, we will consider degrees of nodes together with the comprehensive national strength, forming a competitive network. A competitive model was proposed to describe the competitive mechanism of growth networks [8]. Moreover, the degree distribution of the fitness model is approximate estimated by the competitive model. Therefore, the competitive network for the fitness model has the universality.

In addition, weighted networks can depict competitive advantages of real systems better than un-weighted ones. For example, the weight can denote the amount of paper published cooperatively in the science collaboration network [9,10] or the amount of passenger flow among stations in the transport network [11,12]. In order to explore its property and evolving process deeply, many weighted network models have been proposed[2-4,13-17], and the Barrat-Barthelemy-Vespignani (BBV) model is the one with more influence [15,16]. This model combines the network topology with the dynamical weight to investigate the evolving process. More specifically, an evolving mechanism of preferential attachment based on the vertex (node) strength is proposed.

This paper finds that this competition for links translates into allowing more competitive nodes to overcome the more connected but less competitive ones. This paper also finds that the degree distribution of competitive networks is universal for weighted networks. Owning to the fact that the competitive network is easy to understand and simulate, once the relation between the competitive network and the weighted network is ascertained, it will offer a new perspective for the study on weighted networks. Although there have been a large number of models in complex networks, the problem that which kind of networks has a broader universality has not been discussed in the study of complex networks. This is an important problem. The universality of the competitive network for the fitness model has been obtained[8]. However, do competition networks for weighted networks have the universality?

In this paper we firstly propose a new model that allows us to investigate this competitive aspect of real networks in quantitative terms. We develop the continuum model for this competitive evolving network, and derive a general expression for the connectivity distribution.  In section 3, it is analytically deduced that the BBV weighted network is a special case of the competitive network. In section 4, the theoretical results are verified by discussing the consistency between the two kinds of networks through numerical simulations. Furthermore, the influence of model parameters on degree distributions and clustering coefficients is analyzed as well. Finally, summary and discussions about the universality for other kinds of weighted networks are given in Section 5.

## 2. Competitive networks

In social networks some individuals acquire more social links than others, or on the www some webpages attract considerably more links than others. The rate at which nodes in a network increase their connectivity depends on their competitiveness to compete for links. A competition network is that a node of the network has its own competitiveness; the evolution of the network is not only to relate to the degree of the node, but also to relate to the competitiveness of that node. The competitive network in which the node competitiveness evolving with time is

discussed as follows:

1) *Random growth*: The network starts from initial one with $m_0$ nodes. The arrival process of nodes is a Poisson process having constant rate $\lambda$. When a new node is added to the system at time $t$, this new site is connected to $m$ ($m \le m_0$) previously existing vertices. Each node entering the system is tagged its own $\eta_i$ that we will assume are independent random variables taken from a given distribution $\rho(y)$ characterizing the system's competitiveness, and $a = \int y\rho(y)dy$ are finite.

2) *Preferential attachment*: We assume that the probability that a new node will connect to a node $i$ already present in the network depends on the connectivity $k_i(t)$ and on the competitiveness $\eta_i$ of that node, such that

$$\Pi(k_i(t)) = \frac{k_i(t) + \eta_i}{\sum_i (k_i(t) + \eta_i)}, \qquad (1)$$

$t_i$ denotes the time at which the *ith* node is added to the system. $k_i(t)$ denotes the degree of the $i$th node at time $t$. Assuming that $k_i(t)$ is a continuous real variable, the rate at which $k_i(t)$ changes is expected to be proportional to the degree $k_i(t)$. Consequently, $k_i(t)$ satisfies the dynamical equation

$$\frac{\partial k_i(t)}{\partial t} = m\lambda \frac{k_i(t) + \eta_i}{\sum_i (k_i(t) + \eta_i)} \qquad (2)$$

$N(t)$ represents the total number of nodes that occur by time $t$. By the Poisson process theory, we know $E[N(t)] \approx \lambda t$, thus, for a long time $t$，we have

$$\sum_i (k_i(t) + \eta_i) \approx (2m + a)\lambda t \qquad (3)$$

Substituting Eq.(3) into Eq. (2), we obtain

$$\frac{\partial k_i(t)}{\partial t} = \frac{k_i(t) + \eta_i}{(2 + a/m)t} \qquad t >> t_i \qquad (4)$$

Since $k_i(t_i) = m$，from Eq.(4)，we obtain

$$k_i(t,\eta) = (m + \eta_i)(\frac{t}{t_i})^{\frac{1}{2+a/m}} - \eta_i \qquad t \gg t_i \qquad (5)$$

From Eq.(5)，we obtain

$$P\{k_i(t,\eta) \geq k\} = P\{\frac{k_i(t,\eta) + \eta_i}{m + \eta_i} \geq \frac{k + \eta_i}{m + \eta_i}\} = P\{t_i \leq (\frac{m + \eta_i}{k + \eta_i})^{2+\frac{a}{m}}t\}, \quad t \gg t_i$$

Notice the node arrival process is the Poisson process having rate $\lambda$, therefore the time $t_i$ follows a gamma distribution with parameter $(i,\lambda)$. Thus

$$P\{k_i(t,\eta) < k\} = e^{-\lambda t(\frac{m+\eta_i}{k+\eta_i})^{2+\frac{a}{m}}} \sum_{l=0}^{i-1} \frac{(\lambda t(\frac{m+\eta_i}{k+\eta_i})^{2+\frac{a}{m}})^l}{l!} \qquad t \gg t_i \qquad (6)$$

From Eq. (6), we have

$$P\{k_i(t,\eta) = k\} \approx (2+\frac{a}{m})\frac{\lambda t}{m+\eta_i}(\frac{m+\eta_i}{k+\eta_i})^{3+\frac{a}{m}} \frac{\left[\lambda t(\frac{m+\eta_i}{k+\eta_i})^{2+\frac{a}{m}}\right]^{i-1}}{(i-1)!} e^{-\lambda t(\frac{m+\eta_i}{k+\eta_i})^{2+\frac{a}{m}}} \quad , \quad t \gg t_i \qquad (7)$$

From Eq.(7), we obtain the stationary average degree distribution

$$P(k) \approx (2+\frac{a}{m})\int \frac{1}{m+\eta}(\frac{m+\eta}{k+\eta})^{3+\frac{a}{m}} \rho(\eta)d\eta \qquad (8)$$

If $\rho(\eta) = \delta(\eta - a)$, i.e. all competitiveness are equal, (1) reduces to the initial attractiveness model. Therefore, the initial attractiveness model is a scale-free network with the degree distribution

$$P(k) \approx \frac{2m+a}{m(m+a)}(\frac{m+a}{k+a})^{3+\frac{a}{m}}, \qquad (9)$$

and the degree exponent

$$\gamma = 3 + \frac{a}{m}. \qquad (10)$$

## 3. The universality of the competitive network for the weighted network

In the Internet, it is easy to realize that the introduction of a new connection to a router corresponds to an increase in the traffic handled on the other router's links[16]. Indeed in many technological, large infrastructure and social networks it is commonly believed that a reinforcement of the weights due to the network's growth. In this spirit Barrat et. al. consider here a model for a growing weighted network that takes into account the coupled evolution in time of topology and weights and leaves room for accommodating different mechanisms for the reinforcement of interactions[16].

The definition of the BBV model is based on two coupled mechanisms: the topological growth and the weights' dynamics. Weighted networks are usually described by an adjacency matrix $w_{ij}$ which represents the weight on the edge connecting vertices $i$ and $j$, with $i, j = 1,2,\ldots,N$, where $N$ is the size of the network. We only consider the undirected networks, where the weights matrix are symmetric, that is $w_{ij} = w_{ji}$. The BBV weighted network model is defined as follows [15,16].

(1) Growth: The network starts from an initial $m_0$ nodes connected by edges with assigned weight $w_0$. New nodes arrive the system with a Poisson process having rate $\lambda$. A new node $n$ is added at time $t$. This new site is connected to $m$ ($m \le m_0$) previously existing nodes (i.e., each new node will have initially exactly $m$ edges, all with equal weight $w_0$).

(2) Preferential attachment: The new node $n$ preferentially chooses sites with large strength; i.e., a node $i$ is chosen according to the probability:

$$\Pi_{n \to i} = \frac{s_i}{\sum_j s_j} \tag{11}$$

Where $s_i = \sum_{j \in \Omega(i)} w_{ij}$ is the vertex $i$ strength, the sum runs over the set $\Omega(i)$ of the neighbors of $i$.

(3) Update: The weight of each new edge $(n,i)$ is initially set to a given value $w_0$. The presence of the new edge $(n,i)$ will introduce variations of the existing weights across the network. In particular, we consider the local rearrangements of weights between $i$ and its neighbors $j \in \Omega(i)$ according to the simple rule

$$w_{ij} \rightarrow w_{ij} + \Delta w_{ij} \qquad (12)$$

where $\Delta w_{ij} = \delta \dfrac{w_{ij}}{s_i}$ and $\delta$ is defined as updating coefficient and it is independent on the time $t$.

The changed strength is composed by three parts: the original strength, the new edge weight brought by the new node and the increment of the old edge weight.

When a new vertex $n$ is added to the network, an already present vertex $i$ can be affected in two ways: (i) It is chosen with probability (11) to be connected to $n$; then its connectivity increases by 1, and its strength by $w_0 + \delta$. (ii) One of its neighbors $j \in \Omega(i)$ is chosen to be connected to $n$. Then the connectivity of $i$ is not modified but $w_{ij}$ is increased according to the rule Eq. (12), and thus $s_i$ is increased by $\delta \dfrac{w_{ij}}{s_i}$. This dynamical process modulated by the respective occurrence probabilities $\dfrac{s_i(t)}{\sum\limits_l s_l(t)}$ and $\dfrac{s_j(t)}{\sum\limits_l s_l(t)}$ is thus described by the following evolution equations for $s_i(t)$ and $k_i(t)$:

$$\frac{ds_i}{dt} = m(w_0 + 2\delta) \frac{s_i}{\sum\limits_j s_j} \qquad (13)$$

$$\frac{dk_i}{dt} = m \frac{s_i}{\sum\limits_j s_j} , \qquad (14)$$

Substituting Eq. (14) into Eq. (13) yields:

$$\frac{ds_i}{dt} = (w_0 + 2\delta) \frac{dk_i}{dt}$$

Since node $i$ arrives at the network by time $t_i$, we have $k_i(t_i) = m$ and $s_i(t_i) = mw_0$, then the above equation is integrated from $t_i$ to $t$

$$s_i = (w_0 + 2\delta)k_i - 2\delta m \ , \tag{15}$$

And probability (11) is modified as:

$$\Pi_{n \to i} = \frac{k_i - \dfrac{2\delta}{(w_0 + 2\delta)} m}{\sum\limits_j \left[ k_j - \dfrac{2\delta}{(w_0 + 2\delta)} m \right]} \tag{16}$$

By comparing probability (16) and probability (1) with the distribution $\rho(y) = \delta(y + \dfrac{2\delta m}{w_0 + 2\delta})$, where $\delta(\ )$ denotes the $\delta$ – function, it can be inferred that only if

$$a = -\frac{2\delta}{w_0 + 2\delta} m \ , \tag{17}$$

The probability of the preferential attachment can be modified as $\Pi_{n \to i} = \dfrac{k_i + a}{\sum\limits_j (k_j + a)}$, which is in accord with that of the initial attractiveness model. That is to say, if the updating coefficient $\delta$ is a constant, the corresponding competitiveness $a$ will be a constant too, which verifies the universality on the competitive network for the weighted network.

Moreover, from Eq.(10), the degree distribution of the BBV weighted network behaves as $P(k) \propto k^{-\gamma}$ where

$$\gamma = 3 + \frac{a}{m} = \frac{4\delta + 3w_0}{2\delta + w_0} = 2 + \frac{w_0}{2\delta + w_0} . \tag{18}$$

Therefore, the degree distribution of the weighted network can be obtained directly from the results of the competitive network without discussing the relation between the node degree, strength and the time it arrives.

According to Eq. (15) and Eq. (6), we have:

$$P\{s_i < x\} = P\{k_i < \frac{x + 2\delta m}{w_0 + 2\delta}\} = e^{-\lambda t (\frac{m w_0}{x})^{2 + \frac{a}{m}}} \sum_{l=0}^{i-1} \frac{1}{l!} (\lambda t (\frac{m w_0}{x})^{2 + \frac{a}{m}})^l \tag{19}$$

The density function of $s_i$ is:

$$f_{s_i}(x) \approx (2+\frac{a}{m})\frac{(w_0 m)^{2+\frac{a}{m}}\lambda t}{x^{3+\frac{a}{m}}}e^{-\lambda t(\frac{mw_0}{x})^{2+\frac{a}{m}}}\frac{1}{(i-1)!}\left[\lambda t(\frac{mw_0}{x})^{2+\frac{a}{m}}\right]^{i-1} \tag{20}$$

Then the density function of the stationary average node strength distribution can be deduced from Eq. (20):

$$f(x) \approx (2+\frac{a}{m})\frac{(w_0 m)^{2+\frac{a}{m}}}{x^{3+\frac{a}{m}}}$$

It can be seen by instituting Eq. (17) into the equation above that the density function of the stationary average node strength distribution of the BBV weighted network behaves as power-law function, and the exponent is as same as the degree distribution like $\gamma = 2+\frac{w_0}{2\delta+w_0}$. The density function is

$$f(x) = (\gamma-1)(w_0 m)^{\gamma-1}\frac{1}{x^{\gamma}} \tag{21}$$

## 4. Numerical simulation

The degree distribution and the clustering coefficient are two basic statistics reflecting the topological structure of networks. The former refers to the probability of a certain node to have degree $k$ and the latter refers to the ratio between the number of existing and all possible edges among the neighbors of a certain node. It is known from the previous analytical study that the main factor influencing statistics of the competitive model as $\rho(x)=\delta(x-a)$, is the competitiveness $a$ (it is supposed here that all the nodes have the same competitiveness, i.e., the model is reduced to the initial attractiveness model) and that of the weighted network are the updating coefficient $\delta$ and the initial weight $w_0$. In this section, numerical simulations are carried out to verify the theoretical results above. We find the influence of these factors on the degree distribution and the clustering coefficient, and compare the results of the two networks.

**4.1 The comparison of degree distributions**

Firstly, the influence of parameters on the degree distribution of the competitive network and the weighted network is observed. We take $N = 10000$，$m_0 = 5$，$m = 4$ for both networks. The updating coefficient $\delta$ of the weighted network equals to 0.1, 0.3, 0.5, 0.8, 1 and 1.5, the corresponding the competitiveness $a$ equals to -2/3, -3/2, -2, -32/13, -8/3 and -3 respectively according to Eq. (17). The comparisons are shown in Fig. 1.

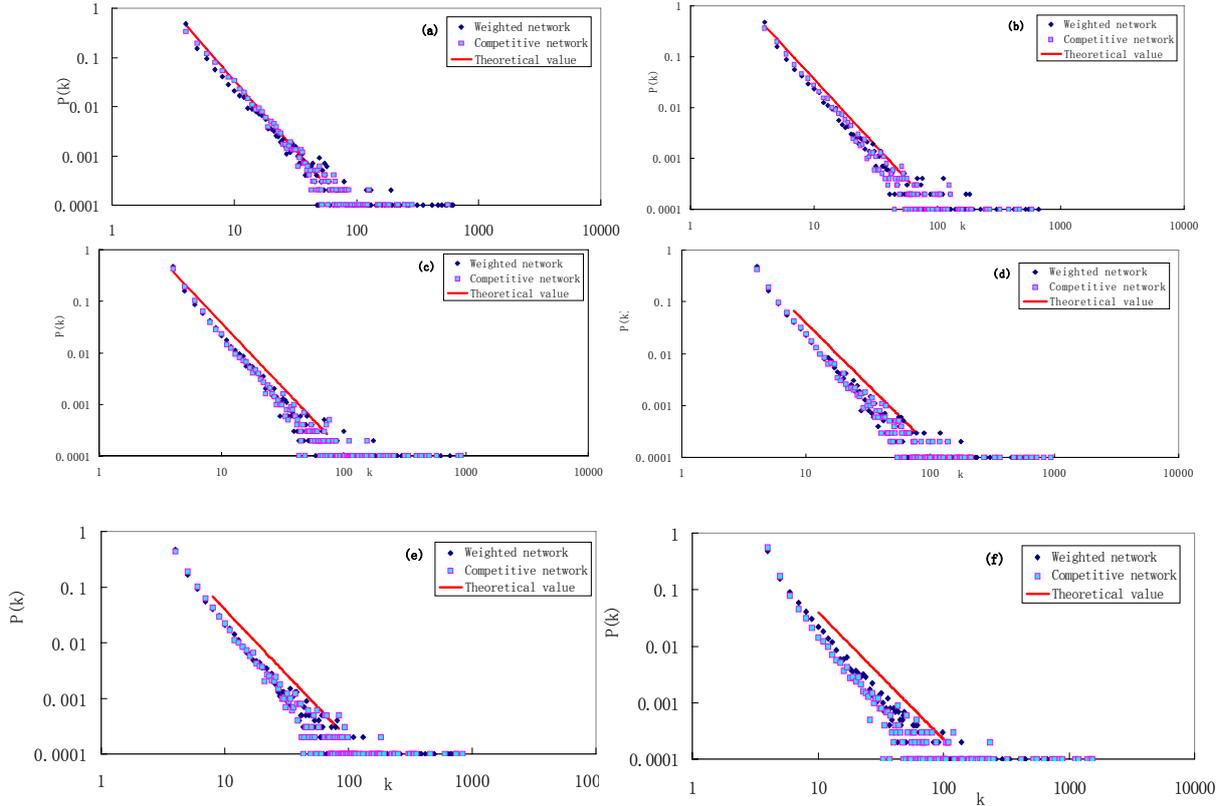

**Fig. 1**. The comparisons of degree distributions of the weighted network and the competitive model. The red lines stand for the theoretical result according to Eq. (18). The sub-plots (a) to (f) correspond to different simulations with $\delta$ = 0.1, 0.3, 0.5, 0.8, 1 and 1.5, $a$ = -2/3, -3/2, -2, -32/13, -8/3 and -3, correspondingly, theoretical value of degree distribution exponent $\gamma$ = 17/6, 21/8, 5/2, 31/13, 7/3 and 9/4, respectively.

The similarity of degree distributions between the two networks is clearly shown in Fig. 1. The slope of main body keeps the same with each other and with the theoretical result. The validity of the analysis in the above section is verified.

### 4.2 The comparison of clustering coefficients

Secondly, the influence of parameters on clustering coefficients of both networks is observed. We take $m_0 = 5$, $m = 4$, $\delta = w_0 = 1$ and $a = -8/3$. The clustering coefficient is calculated along with the growth of networks and the results are shown in Fig. 2.

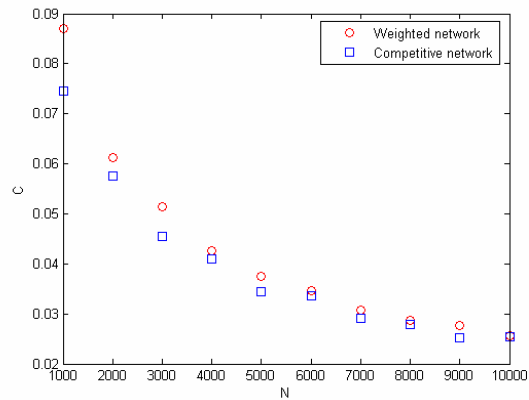

**Fig.2**. The comparisons of clustering coefficients of the weighted network (red circle) and the competitive model (blue square) with different network size $N$.

It is clearly shown in Fig. 2 that the clustering coefficient of the weighted network and that of the competitive network are the same for sufficiently large $N$. At the same time, they show the same tendency that the clustering coefficient gets smaller when the network grows, i.e. the larger the network is, the lower the clustering effect is.

It is found that $\delta$ and $w_0$ have significant impact on the clustering coefficient of the weighted network. We have $N = 10000$, $m_0 = 5$, $m = 4$. The values of $\delta$ and $w_0$ are assigned from 0.1 to 100 respectively. When one of $\delta$ and $w_0$ changes, the other one is fixed to 1. The clustering coefficients are shown in Fig. 3.

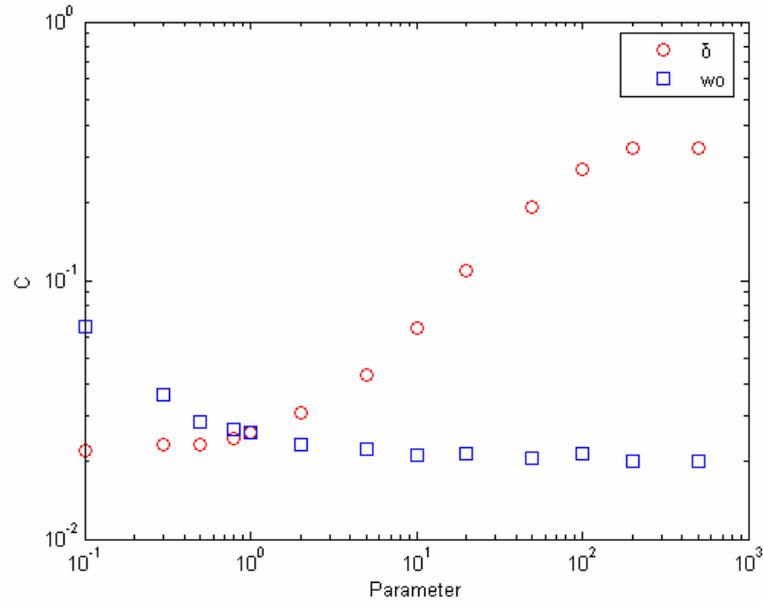

**Fig.3**. The influence of parameter $\delta$ and $w_0$ on the clustering coefficient in the weighted network.

Figure 3 exhibits an interesting shape of the relation between the parameter and the clustering coefficient in a log-log plot. The influence of the values of $\delta$ on the clustering coefficient and that of the value of $w_0$ on the clustering coefficient is symmetric. Since it can be easily found that $\delta$ and $w_0$ is symmetric in Eq.(17), i.e., $a = -\dfrac{2\delta}{w_0 + 2\delta} m = -\dfrac{2(\delta / w_0)}{1 + 2(\delta / w_0)} m$ .

In the evolving process of the weighted network, $m$ edges will be brought into the network by a new node. The impact of parameter $m$ on degree distribution and clustering coefficient is discussed. Let $N = 10000$，$\delta = 1$，$w_0 = 1$，$m_0 = 10$, the simulations are performed as $m$ decreases discretely from 9 to 2. Furthermore, values of the degree distribution and the clustering coefficient $C$ are calculated and the results are shown in Fig. 4 and Fig. 5, respectively.

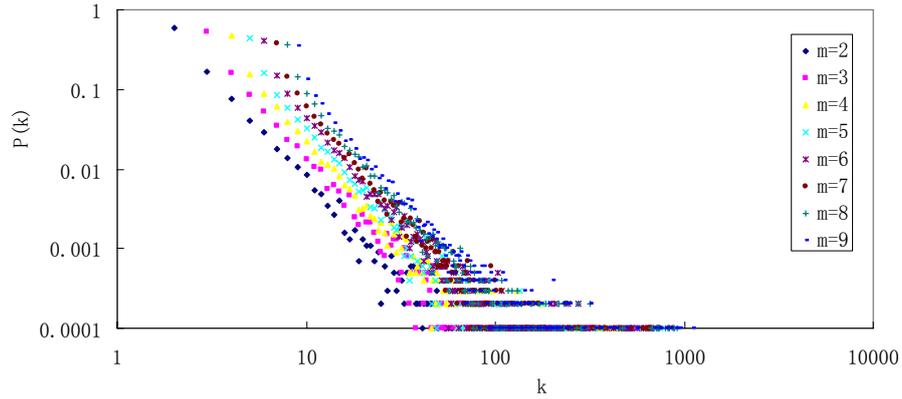

**Fig. 4**. The impact of the number of edges that a new node brings into a weighted network on the degree distribution.

From Fig.4, we can see that the distribution exponent of the weighted network is independent of $m$ , in agreement with the theoretical results, i.e., Eq. (18).

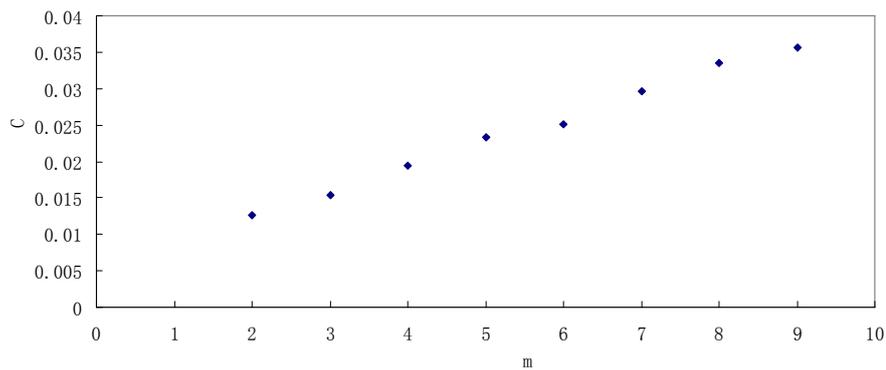

**Fig. 5**. The impact of the number of edges that a new node brings into a weighted network on the clustering coefficient.

Obviously, when a new node is added to the system, it brings $m$ edges, and the connectivity of the network will be more evident, leading to greater clustering coefficient. However, the number of new edges is relatively little impact on the degree exponent of the weighted network.

Finally, we discuss the influence of the competitiveness $a$ on the clustering coefficient of the competitive network. Similarly, let $N = 10000$ , $m_0 = 5$ , $m = 4$ . The simulation results are shown in Fig. 6.

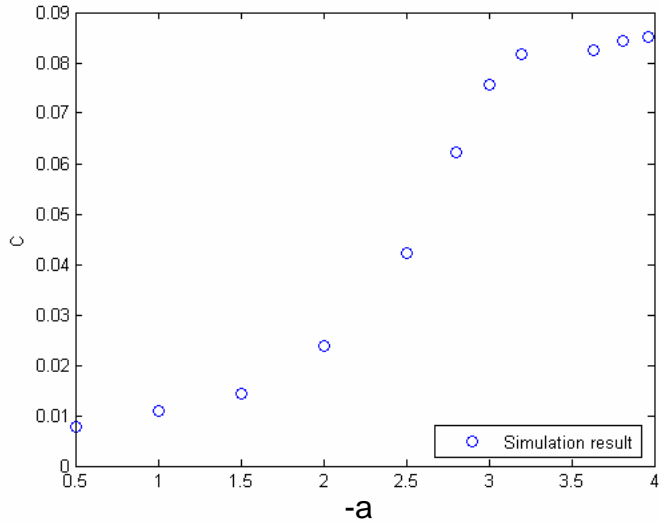

Fig. 6. The impact of the competitiveness on the clustering coefficient. The horizontal ordinate is the opposite number of value $-a$ calculated by Eq. (17).

## 5. Summary and further discussion

The competitive model and the weighted network are two kinds of evolving models with theoretical and realistic significance. The evolving mechanism of them is discussed in this paper. On one hand, the relationship between the competitiveness in the competitive model and updating coefficient, initial weight in the weighted network is given through theoretically analyzing the relation between the preferential attachment probabilities of the two networks. Such analysis reveals the universality on the competitive network for the weighted network, i.e. the weighted network is a special case of the competitive model. On the other hand, numerical simulations are performed to investigate the impact of network parameters, including updating coefficients, competitiveness, initial number of edges, network size and especially the initial weight which is neglected in previous researches on topological property such as degree distribution and clustering coefficient. The simulation results verify the validity of theoretical analysis.

The weighted network model can be divided into two categories according to the cases that weight is assigned with fixed or variable values. The former includes the weighted scale-free (WSF) model[13,14] and the Zheng-Trimper-Zheng-Hui (ZTZH) model[22], and the latter includes the BBV model [15,16] and the Dorogovtsev-Mendes(DM)[17], etc. Most of these evolving models are improved from the BA model. The topological structure of the WSF model or the ZTZH model completely agrees with that of the BA model, i.e. the edge weight is fixed as a constant when the link is established. Therefore, the model can be seen as a special case of the competitive model with the distribution $\rho(x) = \delta(x)$. The probability of preferential attachment in the BBV model depends on the node strength. A newly arrival node brings new edges, leading to the cascading variation on weight and strength. By contrast, the probability of preferential attachment in the DM model is positively correlated with the edge weight. New nodes are brought in by the increase of edge weight, finally giving rise to the network growth. The probability of preferential attachment in the DM model is completely the same as it in the BA model if the increment of edge weight is 0. Since the increment of the edge weight is a constant in the DM model, its preferential attachment mechanism is similar to that of the BBV model. In consideration of the relation of the edge weight and the node strength, the advantage of simple structures and widespread use, there is certain universality to take the BBV model as an example to investigate the relation between the competitive model and the weighted network.

In view of its potentiality and significance to describe real networks, the study on weighted networks has attracted much interest from many scholars [18-21,27], and most of the fruits are about the improvement of an evolving mechanism. Many other aspects such as structural property, dynamical feature and practical application are also worth considering. A deep understanding of the relationship between the competitive model and the weighted network is really helpful to investigate their property and represent the reality.

Web community can be seen as a set of pages that are created by users with similar interests or topics [23]. It can offer some references for the Web community management if we assign every page in WWW with a competitive factor according to the extent of people's interest and divide the community according to such interest. An enterprise business network is set up where a node represents an enterprise and an edge denotes the trade contact between them. This is a competitive network and the competitive factor of nodes indicates the competitive power of the enterprises. Wan and Sun [25] investigates the enterprise collaboration network. Barrat et al [26] regard the world-wide airport network as a weighted network. An empirical research about bus system is studied by Di et. al., which is a weighted network[27].　Yang et. al. constructed the annual competitive relationship complex network models among enterprises according to data of enterprises and products among software industry from the year 2002 to 2006 in Guangdong Province, China[24]. Yang et. al. proposed a product-competition network of industrial towns[28]. Through the analysis of the real industry data, they verified the product-competition model. Further research on the application of competitive networks and weighted networks is a significant problem.

**Acknowledgement**

This paper is supported by the National Natural Science Foundation of China (Grant No.70871082) and the Shanghai First-class Academic Discipline Project (Grant No. S1201YLXK). The authors thank referees' useful comments.

**References**

[1] Albert, R. and Barabasi, A. L. Statistical mechanics of complex networks, *Rev. Mod. Phys.* 74 (2002) 47-97.

[2] Li M H, Wang D H, Fan Y, Di Z R, and Wu J S. Modelling weighted networks using connection count, *New Journal of Physics 8 (2006) 72.*

[3] Pan Z F, Wang X F. A weighted scale-free network model with large-scale tunable clustering, *Acta Phys. Sin., 2006,55(8):4058-4064.* (in Chinese)

[4] Boccaletti, S., Latora, V., Moreno, Y., Chavez, M. and Hwang, D. U., Complex networks: Structure and dynamics, *Phys. Reports* 424 (2006) 175-308.

[5] Newman, M., The physics of networks, *Phys. Today* 11 (2008) 33–38.

[6] Bianconi, G. and Barabasi, A. L., Bose-Einstein Condensation in Complex Networks, *Phys. Rev. Lett.* 86 (2001) 5632-5635.

[7] Dorogovtev, S. N., Mendes, J. F. F. and Samukhin, A. N., Structure of growing networks with preferential linking, *Phys. Rev. Lett.* 85 (2000) 4633-4636.

[8] Guo, J. L., Poisson growing competition networks and the fitness model, *Mathematics in practice and theory*, 40 (2010) 175-182. (in Chinese)

[9] Newman, M. E. J., The structure of scientific collaboration networks, *Proc. Natl. Acad. Sci. USA* 98 (2001) 404-409.

[10] Li M H, Wu J S, Wang D H, Zhou T, Di Z R, Fan Y. Evolving model of weighted networks inspired by scientific. Physica A 375 (2007) 355–364.

[11] Park, K., Lai, Y. C. and Ye, N., Characterization of weighted complex networks, Phys. Rev. E 70 (2004) 026109.

[12] Allman, M. and Paxson, V., On estimating end-to-end network path properties, *ACM SIGCOMM Computer Communication Review* 31(2001) 124.

[13] Yook, S. H., Jeong, H. and Barabasi, A. L., Weighted Evolving Networks, *Phys. Rev. Lett.* 86 (2001) 5835-5838.

[14] Eom, Y. H., Jeon, C., Jeong, H. and Kahng, B., Evolution of weighted scale-free networks in empirical data, *Phys. Rev. E* 77 (2008) 056105.

[15] Barrat, A., Barthelemy, M. and Vespignani, A., Weighted Evolving Networks: Coupling Topology and Weight Dynamics, *Phys. Rev. Lett.* 92 (2004) 228701.

[16] Barrat, A., Barthelemy, M. and Vespignani, A., Modeling the evolution of weighted networks, *Phys. Rev. E* 70 (2004) 066149.

[17] Dorogovtsev, S. N. and Mendes, J. F. F., Minimal models of weighted scale-free networks, arXiv: cond-mat/0408343v2.

[18] Wang, W. X., Wang, B. H., Hu, B., Yan, G. and Ou, Q., General dynamics of topology and traffic on weighted technological networks, *Phys. Rev. Lett.* 94 (2005) 188702.

[19] Ramasco, J. and Gonçalves, B., Transport on weighted networks: when correlations are independent of degree, *Phys. Rev. E.* 76 (2007) 066106.

[20] Serrano, M., Boguná, M. and Vespignani, A., Extracting the multiscale backbone of complex weighted networks. Proc. Natl. Acad. Sci. USA 106 (2009) 6483–6488.

[21] Riccaboni, M. and Schiavo, S., Structure and growth of weighted networks, *New J. Phys* 12 (2010) 023003.

[22] Zheng, D. F., Trimper, S., Zheng, B. and Hui, P. M., Weighted scale-free networks with stochastic weight assignments, *Phys. Rev. E* 67 (2003): 040102.

[23] He, D., Peng, Z. Y. and Mei X. R, Brief Survey of Web Community Management Research, *Journal of Frontiers of Computer Science and Technology* 5(2011) 97-113. (in Chinese)

[24] Hu X, Yang J M, Li D R. The Complex Network Analysis of the Enterprise Competitive Relationships Evolution —Take Software Industry in Guangdong Province as the Example, Soft Science, 2008, 22(6): 52-73. (in Chinese)

[25] Wan, L. J. and Sun, Y. X., The Research on Degree Distribution of Cooperative Networks in Corporation, *Mathematical Theory and Applications* 29(2009) 60-64. (in Chinese)

[26] Barrat, A., Barthélemy, M., Pastor-Satorras, R. and Vespignani, A., The architecture of complex weighted networks, *Proc. Natl. Acad. Sci. USA* 101 (2004) 3747-3752.

[27] Gao L, Zhao J S, Di Z R, Wang D H. Asymmetry between odd and even node weight in complex networks, *Physica A* 376 (2007) 687–691.

[28] Li M N, Yang J M, Tang W L. Research on features of product-competition network based on complex network, Electronic Measurement Technology, 2007, 30(4): 8-11. (in Chinese)